\documentclass[10pt,sigconf,letterpaper]{acmart}
\newcommand{\longn}[0]{{Multicast QUIC}}
\newcommand{\shortn}[0]{{MCQUIC}}
\usepackage{siunitx}
\usepackage[inline]{enumitem}
\usepackage{tikz}
\usepackage{xcolor}
\usepackage{caption}
\usepackage{subcaption}
\usepackage[super]{nth}
\usetikzlibrary{fit}
\usetikzlibrary{shapes}
\usepackage{todonotes}
\usepackage{tabularx}
\usetikzlibrary{matrix,arrows.meta,arrows}
\usetikzlibrary{calc}

\usepackage{listings}     
\usepackage{lstautogobble}  
\usepackage{color}          
\usepackage{zi4}            

\definecolor{bluekeywords}{rgb}{0.13, 0.13, 1}
\definecolor{greencomments}{rgb}{0, 0.5, 0}
\definecolor{redstrings}{rgb}{0.9, 0, 0}
\definecolor{graynumbers}{rgb}{0.5, 0.5, 0.5}

\usepackage{listings}
\setcopyright{none}
\renewcommand\footnotetextcopyrightpermission[1]{} 
\tikzset{
	cross/.pic = {
		\draw[thick,rotate = 45] (-#1,0) -- (#1,0);
		\draw[thick,rotate = 45] (0,-#1) -- (0, #1);
	}
}

\definecolor{myorange}{HTML}{de8e04}
\definecolor{mygreen}{HTML}{029d74}
\definecolor{mypurple}{HTML}{cc78bd}
\definecolor{myblue}{HTML}{0174b2}

\begin{document}

\title{\shortn{}: Multicast and unicast in a single transport protocol}


\author{Louis Navarre}
\affiliation{%
   \institution{UCLouvain}
   \country{Belgium}
}
\email{louis.navarre@uclouvain.be}

\author{Olivier Pereira}
\affiliation{%
   \institution{UCLouvain}
   \country{Belgium}
}
\email{olivier.pereira@uclouvain.be}

\author{Olivier Bonaventure}
\affiliation{%
   \institution{UCLouvain}
   \country{Belgium}
}
\email{olivier.bonaventure@uclouvain.be}

\renewcommand{\shortauthors}{Louis Navarre et al.}

\begin{abstract}
  Multicast enables efficient one-to-many communications. Several applications benefit from its scalability properties, e.g., live-streaming and large-scale software updates. Historically, multicast applications have used specialized transport protocols. The flexibility of the recently standardized QUIC protocol opens the possibility of providing both unicast and multicast services to applications with a single transport protocol.  
  
  We present \shortn{}, an extended version of the QUIC protocol that supports multicast communications. We show how QUIC features and built-in security can be leveraged for multicast transport. We present the design of \shortn{} and implement it in Cloudflare \emph{quiche}. We assess its performance through benchmarks and in emulated networks under realistic scenarios. We also demonstrate \shortn{} in a campus network. 
  By coupling QUIC with our multicast extension, applications can rely on multicast for efficiency with the possibility to fall back on unicast in case of incompatible network conditions. 

\end{abstract}



\maketitle
\pagestyle{plain}

\section{Introduction}
Many Internet users need to receive the same content, e.g., live multimedia broadcast, operating systems and software updates, news distribution, \ldots 
Today, this popular content is mainly distributed by Content Distribution Networks (CDNs). Each of these CDNs manages hundreds or thousands of edge servers that are capable of serving the same content, using TCP and recently QUIC. Unfortunately, the same content often passes over a given Internet link multiple times. 

This architecture consumes a large amount of resources (edge servers, bandwidth, \ldots) to deliver content at a large scale. IP Multicast, initially proposed by Deering in 1990~\cite{deering1990multicast}, has the potential to deliver the same content with fewer resources. With IP Multicast, networks can dynamically create multicast forwarding trees rooted at a sender and with all the receivers as leaves. The intermediate routers that compose those trees automatically duplicate the multicast packets received from the upstream node of a given tree to the different downstream nodes. Two main variants of IP Multicast have been developed within the IETF: Any Source Multicast (ASM) and Single Source Multicast (SSM). ASM was the initial multicast architecture, but it did not scale well in large networks~\cite{diot2000deployment} and interdomain ASM has been deprecated by the IETF \cite{rfc8815}. Internet Service Providers (ISPs) prefer the SSM architecture and often use it to deploy IPTV services~\cite{maisonneuve2009overview}. SSM enables one identified sender to multicast packets towards a set of receivers that have joined the multicast tree rooted at this particular sender. This multicast tree is identified by the IP address of the server and a group address that all interested receivers have joined. SSM is used to support IPTV in ISP networks~\cite{mellia2010measurement} and virtualized workloads in datacenters~\cite{komolafe2017ip}, distribute files in enterprise networks~\cite{microsoft:2022} and provide financial services~\cite{cisco}. This paper focuses on the SSM architecture.

Applications cannot use IP Multicast directly. They must use a multicast-capable transport protocol. Today, most multicast applications rely on UDP or run another protocol such as RTP on top of UDP. Researchers proposed a variety of multicast transport protocols~\cite{obraczka1998multicast,gemmell2003pgm}, but none of these approaches became widely deployed. During the last years, based on an initial prototype developed by Google~\cite{roskind2012quick,langley2017quic}, the IETF standardized the QUIC protocol~\cite{rfc9000}. QUIC was initially designed to support HTTP/3~\cite{rfc9114}. It combines the reliability and congestion control mechanisms of modern TCP implementations with the security features of TLS 1.3~\cite{rfc8446}. QUIC supports multiple streams and can carry both reliable data and datagrams~\cite{rfc9221}. Most of the packets exchanged by QUIC peers are encrypted, apart from a very short header. More importantly, QUIC packets carry frames and it is relatively easy to extend the protocol by adding new frames types. Extensions such as Multipath QUIC~\cite{ietf-quic-multipath-04} leverage this extensibility. QUIC is largely deployed by CDNs~\cite{zirngibl2021s}. QUIC is usually implemented as a library which can be directly included in the application~\cite{jaeger2023quic,msquic}. Since QUIC runs above UDP, it can use IP Multicast destination addresses, enabling it to become a transport protocol standardized by the IETF which can support both unicast and multicast.
This property enables reconsidering the use of multicast. Instead of entirely relying on multicast for one-to-many communication, this modularity shifts into a \emph{multicast as a service} paradigm. Multicast could be used when possible for better use of resources, but the communication could easily fall back on a unicast session seamlessly for the client. This is important in today's Internet where different access networks support different types of services with different levels of multicast deployment.

This paper presents \longn{} (\shortn{}), a multicast extension of the QUIC transport protocol offering group key security, source authentication and partial reliability. \shortn{} relies on Multipath QUIC~\cite{ietf-quic-multipath-04} to implement a multicast forwarding path transparently for the clients. We provide an implementation of \shortn{}, evaluate it with benchmarks and deploy it in our campus network. We also use emulated networks to validate \shortn{} in lossy scenarios. We extended Cloudflare's \emph{quiche}~\cite{quiche} with full backward compatibility with unicast QUIC as defined in RFC9000~\cite{rfc9000}. The implementation of \shortn{} required $\sim$7000 source lines of code (SLoC) for the core behavior and $\sim$5000 SLoC for tests. We will publicly release our source code upon upcoming publications. This work does not raise any ethical issue.

This paper is organized as follows. Section~\ref{sec:quic} summarizes the main features of QUIC~\cite{rfc9000} and Multipath QUIC~\cite{ietf-quic-multipath-04} that we use in \shortn{}. 
We detail the design of \shortn{} in Section~\ref{sec:design} and evaluate it through benchmarks in Section~\ref{sec:benchmarks}. Section~\ref{sec:evaluation} shows how we deployed our implementation in a real multicast network used in our campus. It further explores the reliability of \shortn{} with losses in emulated networks. 
We conclude this paper in Section~\ref{sec:conclusion}.

Multicast extensions for QUIC have already been discussed within the IETF~\cite{pardue-quic-http-mcast-11, jholland-quic-multicast-02}. Even if this paper partially builds upon these drafts for the design of \shortn{}, it suggests different mechanisms for reliability, source authentication and modularity with the unicast session between the server and each client. Additionally, the design presented in this paper is evaluated in a real implementation. This is further discussed in Section~\ref{sec:design-ietf}.

\section{QUIC Background}\label{sec:quic}
QUIC~\cite{rfc9000} is a connection-oriented protocol built on top of UDP. The majority of a QUIC packet is encrypted to prevent middleboxes from acting on it. TLS 1.3 is embedded in QUIC to provide a fast and secure session handshake. 
A QUIC connection starts with a handshake where TLS session keys are computed for the communication. In contrast to the 4-tuple used in TCP, a QUIC connection is identified with a source and destination Connection ID (CID), generated at random by the end-hosts. Each QUIC packet is identified with a monotonically increasing packet number (PN). Retransmitted data is sent in new QUIC packets with increased packet numbers. These packets can be considered as \emph{frame} containers, which carry control and data information. Thanks to this architecture, it becomes easy to extend the protocol by defining new such frames. QUIC supports reliable, ordered data stream multiplexing through the \texttt{STREAM} frame, and unreliable communication with the \texttt{DATAGRAM} frame. Each stream is assigned a Stream ID (SID) carried in the \texttt{STREAM} frame. The SID is encoded on 62 bits, meaning that more than $4.6\times 10^{18}$ streams can be used in a single QUIC connection.

Multipath QUIC~\cite{de2017multipath} extends the QUIC protocol by allowing to simultaneously use multiple paths  within a single QUIC connection. Like Multipath TCP \cite{rfc8684}, Multipath QUIC enables to either improve the transfer bandwidth or the connection reliability~\cite{de2017multipath}. Multipath QUIC associates each path with a distinct CID, exchanged between the end-hosts using the \texttt{NEW\_CONNECTION\_ID} frame. A new path can only be used once it has been probed by a host and acknowledged by its peer with the \texttt{PATH\_CHALLENGE} and \texttt{PATH\_RESPONSE} frames. The IETF is finalizing the standardization of this extension~\cite{ietf-quic-multipath-04}.
Several open source implementations\footnote{See \url{https://github.com/quicwg/multipath/wiki/QUIC-Implementations-with-multipath-support}.} of QUIC already support multipath~\cite{picoquic} or discuss its integration~\cite{quiche-multipath}. 
Multipath QUIC support is negotiated during the handshake with the \texttt{enable\_multipath} transport parameter.

\section{\shortn{} design}\label{sec:design}
This section describes the design of \shortn{}.
We first present how a server advertises a multicast channel to its client in Section~\ref{sec:design-channel}. Section~\ref{sec:multipath} shows how we leverage Multipath QUIC to transparently transmit multicast data to the clients, enabling unicast fall-back seamlessly. Section~\ref{sec:reliability} introduces the multicast reliability mechanism used in \shortn{} to enable for packet recovery by relaxing the full-reliability constraint and introducing a data expiration timer defined by the application. 
Finally, we discuss extensions to support source authentication in Section~\ref{sec:authentication}. Such mechanism can be used by an application to ensure that clients receive data from the expected source.
Due to space limitations, the \shortn{} frames are detailed in Appendix~\ref{sec:frames} (\tablename~\ref{table:frame-announce} to~\ref{table:frame-auth}).

\subsection{Multicast channel announcement}\label{sec:design-channel}
Any \shortn{} communication starts with a standard QUIC handshake~\cite{rfc9000}. During this handshake, both endpoints advertise their local support of multicast with two new transport parameters, respectively for server and client support (\texttt{mc\_server\_params} and \texttt{mc\_client\_params}).
Multicast communication is only possible if both the client and the server support multicast. Multipath QUIC must also be enabled.

\begin{figure}
\centering
\begin{minipage}{.48\textwidth}
  \centering
  \resizebox{.95\linewidth}{!}{\begin{tikzpicture}[>=latex]
	\draw[->, very thick] (-3,0) -- (-3,-4.2);
	\draw[->, very thick] (0,0) -- (0,-4.2);
	\draw[->, very thick]  (3,0) -- (3,-4.2);
	\node at (-3,.3) {UC server};
	\node at (0,.3) {Client};
	\node at (3,.3) {MC source};
	\draw[->,very thick, mygreen] (0,-0.5) -- node[midway,above,sloped] {OK(MC client)} (-3, -0.75);
	\draw[->,very thick, mygreen] (-3,-0.75) -- node[midway,below,sloped] {OK(MC server)} (-0, -1);
	\draw[->,very thick, mygreen] (-3,-1.75) -- node[midway,above,sloped] {MC\_ANNOUNCE} (-0, -2);
	\draw[->,very thick, mygreen] (0,-2) -- node[midway,below,sloped] {MC\_STATE(Join)} (-3, -2.25);
	
	\draw[->,very thick, mygreen] (-3,-3) -- node[midway,above,sloped] {MC\_KEY} (-0, -3.25);
	\draw[->,densely dotted,myblue,very thick] (-3, -3.5) -- (3, -3.5);
	\draw[->,very thick, mygreen] (0,-3.25) -- node[midway,below,sloped] {MC\_STATE(List.)} (-3, -3.5);
	
	\draw[->,very thick, densely dashed, myorange] (3,-3.5) -- node[midway,below,sloped] {\texttt{STREAM($y$)}, $x$} (0, -3.75);
	
\end{tikzpicture}}
  \captionof{figure}{Multicast channel announcement.}
  \label{fig:design:announcement}
\end{minipage}
\begin{minipage}{.48\textwidth}
  \centering
  \resizebox{.95\linewidth}{!}{\begin{tikzpicture}[>=latex]
	\draw[->, very thick] (-3,0) -- (-3,-4);
	\draw[->, very thick] (0,0) -- (0,-4);
	\draw[->, very thick]  (3,0) -- (3,-4);
	\node at (-3,.3) {UC server};
	\node at (0,.3) {Client};
	\node at (3,.3) {MC source};
	\draw[->,very thick, myorange,densely dashed] (3,-0.5) -- node[midway,above,sloped] {\texttt{STREAM(1)}, $x$} (0, -1);
	\draw[-,very thick,red,densely dashed] (3,-1.25) -- node[pos=1,above,sloped] {\texttt{STREAM(5)}, $x + 1$} (1.5, -1.5);
	\path (1.5,-1.5) pic[red] {cross=4pt};
	\draw[->,very thick, myorange, densely dashed] (3,-2) -- node[midway,above,sloped] {\texttt{STREAM(9)}, $x + 2$} (0, -2.5);
	\draw[->, densely dotted,myblue,very thick] (-3, -3) -- (3, -3);
	\draw[->,very thick,mygreen] (0,-2.5) -- node[midway,above,sloped] {\texttt{MC\_NACK($x+1$)}} (-3, -3);
	
	\draw[->,very thick, densely dashed, myorange] (3,-3) -- node[midway,below,sloped] {\texttt{REPAIR($x..x+2$)}} (0, -3.5);
	
\end{tikzpicture}}
  \captionof{figure}{Multicast reliability mechanism with negative acknowledgments and Forward Erasure Correction.}
  \label{fig:design:reliability}
\end{minipage}%
\vspace{-1em}
\end{figure}

After the QUIC handshake, the unicast server potentially announces the multicast channel information (\figurename~\ref{fig:design:announcement}) with an \texttt{MC\_ANNOUNCE} frame (\tablename~\ref{table:frame-announce}). This frame contains the Channel ID (the multicast equivalent of the Connection ID), the multicast source and group IP addresses, and the UDP destination port.
Group management actions are performed with the \texttt{MC\_STATE} frame (\tablename~\ref{table:frame-state}).
To join the multicast channel, a client sends this frame with the \texttt{JOIN} action. The server responds with an \texttt{MC\_KEY} frame (\tablename~\ref{table:frame-key}) containing the multicast session key used to decrypt packets sent on the multicast channel. The client is now considered a member of the \shortn{} group.

The unicast session between the client and the server remains open during all communication, even if the client is in the multicast group.
During this multicast handshake, the client potentially receive application data through this unicast session.

\subsection{Multicast leveraging Multipath}\label{sec:multipath}

Multipath QUIC (MPQUIC) is being standardized within the IETF~\cite{ietf-quic-multipath-04}. Thanks to this extension, it becomes possible to leverage multiple paths in a single QUIC connection. %

\textbf{Overview.} Introducing the \emph{multicast as a service} paradigm, a client can transparently switch from the multicast channel to the unicast session if minimum conditions are not met to stay in the multicast group (e.g., too many packet losses or change in the support of IP multicast in the network). 
\shortn{} relies on MPQUIC to create the multicast data forwarding channel as a second path.

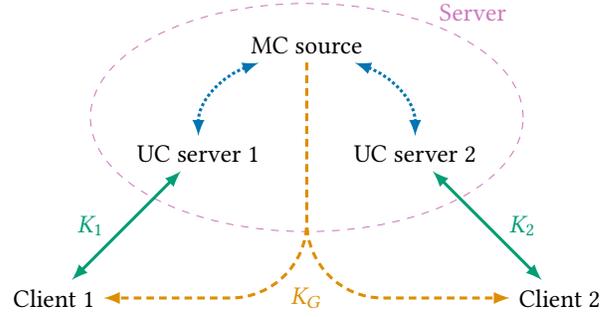
\begin{figure}
\centering
\begin{minipage}{.48\textwidth}
  \centering
  \resizebox{.95\linewidth}{!}{\begin{tikzpicture}[>=latex]

\coordinate (bottom left) at (-2, -2);
		\coordinate (top right) at (2, 0);
		\node[draw=mypurple, text=mypurple, ellipse, dashed,fit={(bottom left) (top right)}, align=right, text height=1ex] {};
		\node[text=mypurple] (server-total) at (2.3, 0.45) {Server};

		\node[] (mc-source) at (0, 0) {MC source};
		\node[] (uc-server-1) at (-1.5, -1.5) {UC server 1};
		\node[] (uc-server-2) at (1.5, -1.5) {UC server 2};
		
		\node[] (client-1) at (-3.5, -3.5) {Client 1};
		\node[] (client-2) at (3.5, -3.5) {Client 2};
		
		\draw[<->, mygreen, very thick] (client-1) to node[left=0.5em]{$K_1$} (uc-server-1);
		\draw[<->, mygreen, very thick] (client-2) to node[right=0.5em]{$K_2$} (uc-server-2);
		
		\draw[<->, densely dotted, very thick,color=myblue] (mc-source) to [out=200,in=90] (uc-server-1);
		\draw[<->, densely dotted, very thick,color=myblue] (mc-source) to [out=340,in=90] (uc-server-2);
		
		\draw[rounded corners=30pt, ->,color=myorange, very thick, densely dashed] (mc-source.south) |- (client-1.east) {};
		\draw[rounded corners=30pt, ->,myorange, very thick, densely dashed] (mc-source.south) |- (client-2.west) {};
		\node[text=myorange] (k-g) at (0, -3.5) {$K_G$};

\end{tikzpicture}}
  \captionof{figure}{\shortn{} leverages Multipath QUIC to create the multicast data channel. Both clients have a unicast session with the server (\emph{UC server}) encrypted and authenticated with their unicast session keys ($K_1$ and $K_2$). Clients listen to the multicast group address announced by the server as second path. They decrypt and authenticate the data using the multicast group session key $K_G$. The server-side unicast connections exchange control information with the multicast source (\emph{MC Source}).}
  \label{fig:design:multipath}
\end{minipage}
\hfill
\begin{minipage}{.48\textwidth}
  \begin{lstlisting}[caption={Scheduling of the repair symbols of the FEC recovery mechanism.},label=list:fec]
fn recv_nack(ranges, p_max):
	n = nb_reliable_lost(ranges)
	n_repairs = nb_sent_repair_since(p_max)
	nb_to_send_local = max(0, n - n_repairs)
	nb_to_send = max(nb_to_send, n)

fn should_send_repair():
	return nb_to_send > 0 and nb_to_send < max_nb_repair
\end{lstlisting}
\end{minipage}%
\vspace{-1em}
\end{figure}

\figurename~\ref{fig:design:multipath} shows two clients listening to the multicast channel.
The first path is the unicast session between the client and the server. The unicast session keys ($K_1$ and $K_2$) are specific to this path. The second path is encrypted using a multicast session key ($K_G$) that is derived by the multicast source, and forwarded to the clients on the unicast paths using the \texttt{MC\_KEY} frame. 
From the client's perspective, the multicast channel is hence a second path using a different cryptographic context.  

\textbf{Multicast path.} For $n$ clients, the \shortn{} server (\emph{Server} in \figurename~\ref{fig:design:multipath}) manages a set of $n+1$ QUIC connections. For each new client, the \emph{Server} start a new unicast QUIC connection (\emph{UC server}). 
The additional connection is the multicast source (\emph{MC source}). When a client joins the \longn{} channel, it creates a new path towards its \emph{UC server}. 
The four-tuple associated to this path uses the multicast group address and port advertised in the \texttt{MC\_ANNOUNCE} frame.
The multicast path is unidirectional and only the \emph{MC source} sends data on this path.

\textbf{Group session key.} A secure multicast communication requires that each member of the group has access to the common group session key, $K_G$~\cite{hardjono2000ip}. 
This key is different from the unicast session keys ($K_1$ and $K_2$ in \figurename~\ref{fig:design:multipath}) to ensure the integrity of the unicast sessions. 
$K_G$ is derived by the \emph{MC source}.
Once a client joins the group, it receives the secrets used to derive $K_G$ through the \texttt{MC\_KEY} frame sent by the unicast server. When the client receives a packet on the multicast path, it uses $K_G$ to decrypt it.
Group key management in multicast is an important property to ensure backward and forward-secrecy properties of communications. Although this remains an open research problem~\cite{hardjono2000ip,wallner1999key}, we consider it as orthogonal to the design of \shortn{} and is not considered in this paper. Indeed, a group key management algorithm could be integrated on top of this design by leveraging \texttt{MC\_KEY} frames and the multicast channel.

\textbf{Multicast as a service.} 
Since the multicast channel is considered as a second path by the client, it
may seamlessly switch from multicast to unicast and still receive application data through the unicast path without interruption.
This is possible thanks to the features of Multipath QUIC.
A client leaves the channel by sending an \texttt{MC\_STATE} frame with the \texttt{LEAVE} action to its \emph{UC server}. 
We detail in Section~\ref{sec:reliability} how the clients use the unicast channels to return acknowledgments to the source to provide partial reliability.



\subsection{Reliability}\label{sec:reliability}
Ensuring (partial) reliability is essential in multicast communications. As noted by the authors of PGM~\cite{gemmell2003pgm}, for a group of 1,000,000 clients with an independent loss probability of \SI{0.01}{\percent}, the probability that all clients receive a given packet is $10^{-43}$. Most strategies use a system of negative acknowledgment (NACK) from the clients to the source to notify lost packets~\cite{gemmell2003pgm}. Other designs suggest the use of a hierarchical distribution of the clients to provide reliability~\cite{paul1997reliable}. Another technique to recover from packet losses in multicast is Forward Erasure Correction (FEC)~\cite{gemmell2003pgm, byers2002digital, rizzo1997effective, rfc3453}, as a single repair symbol can recover different lost packets on distinct receivers.

\textbf{Overview.} \longn{} relies on a NACK strategy which consumes less bandwidth and resources on the server than positive acknowledgments. To inform the source of lost packets, clients send NACKs on the unicast session.
Because \shortn{} supports Single-Source Multicast (SSM), there is no interaction between receivers to perform NACK aggregation~\cite{desmouceaux2018reliable}. For simplicity reasons, the design does not support hierarchical distribution of the receivers either~\cite{paul1997reliable}.

\shortn{} provides partial reliability by sending Forward Erasure Correction (FEC) repair frames on the multicast path. The unicast path between a client and its \emph{UC source} uses the standard QUIC retransmission mechanisms~\cite{rfc9002}. \shortn{} uses an \emph{expiration timer} (ET) set by the application. This value indicates the time during which the source can send repair frames to recover lost data. Upon expiration, every data sent before this timer are removed from the sending state of the source, and no more repair symbols will be generated to recover them.

\textbf{Negative acknowledgments.} A unicast QUIC end-host sends positive acknowledgments (\texttt{ACK} frames) to its peer with the ranges of packets numbers (PN) correctly received. The peer detects lost packets by inspecting the gaps in these ranges, and retransmits the lost reliable frames (e.g., \texttt{STREAM} frames) in subsequent packets. 
Instead, \shortn{} clients send \texttt{MC\_NACK} frames (\tablename~\ref{table:frame-nack}) to their \emph{UC server}, as presented in \figurename~\ref{fig:design:reliability}. This frame contains the ranges of missing packet numbers and the highest PN (\texttt{p\_max}) the client received at the time it generated the frame. As shown in the \figurename, clients detect a gap in the packet sequence numbers once they receive packets with a higher PN.
Clients do not send \texttt{ACK} frames for packets received on the multicast path, but use the standard reliability mechanism~\cite{rfc9002} on the unicast path.
Similarly to unicast QUIC, the \shortn{} source can regularly send \texttt{PING} frames if no application data must be sent, to trigger \texttt{MC\_NACK} from the clients.

\textbf{Retransmissions using FEC.}
Lost frames sent on the multicast path are not directly retransmitted by the source. Instead, the emitter sends Forward Erasure Correction (FEC) repair packets on the multicast path to all clients. Using these repair symbols with the correctly received packets, a client can recover the frames sent in the lost packets. 
Replacing the QUIC retransmission mechanism with this strategy already shown positive results for unicast communication~\cite{michel2022flec}.
In multicast, a single repair packet can be sufficient to recover different losses on distinct receivers.
We follow the FEC for QUIC loss recovery draft~\cite{michel-quic-fec-00}. Frames to be protected by FEC are mapped into source symbols using the \texttt{SOURCE\_SYMBOL} header. The mapping consists in adding a monotonically increasing source symbol ID (SSID) to each protected frame. Repair symbols are carried inside \texttt{REPAIR} frames. 
In \shortn{}, every \texttt{STREAM} frame sent on the multicast path is mapped into a source symbol.

Listing~\ref{list:fec} presents our FEC scheduler to send repair packets. The source maintains two state variables, \texttt{nb\_to\_send} indicating the number of repair symbols to generate, and \texttt{sent\_repairs} containing the PNs of already generated repair symbols. 
When the source receives an \texttt{MC\_NACK} from a client, it computes the number of packets (containing reliable frames) the client did not receive (Line 2).
Lines 3 and 4 limit the number of repair symbols to send using the feedback from this client. For example, in case the client has a higher delay than others, it may ask for repair packets that were already generated by the source. Line 3 uses \texttt{p\_max} to compute the number of repair symbols the source already generated that the client was not aware of when sending the \texttt{MC\_NACK}. The number of repair symbols to generate for this client is subtracted by this value. The source generates as many repair symbols as indicated by \texttt{nb\_to\_send}. This value is decremented for each generated and sent repair symbol.

\textbf{Partial reliability and expiration timer.}
Ensuring theoretically full reliability in multicast requires that the forwarding follows the bottleneck client. Instead, \shortn{} relaxes this constraint by offering partial reliability controlled by an \emph{expiration timer} (ET) set by the application. \shortn{} adds a new timer triggered every ET. Packets that were sent before the previous ET are considered expired by the source, and removed from the sending state; streams that were open before the last ET are reset. The FEC state is also reset, i.e., expired source symbols are removed from the window and expired repair symbols are removed from \texttt{sent\_repairs}.

The source notifies group members of expired packets/streams/FEC symbols by sending an \texttt{MC\_EXPIRE} frame (\tablename~\ref{table:frame-expire}) at each ET trigger with
\begin{enumerate*}[label=(\roman*)]
    \item the highest expired packet number,
    \item the highest expired stream ID (SID) and
    \item metadata to reset the FEC state.
\end{enumerate*}
Since packet numbers and SSIDs monotonically increase, a receiver can safely remove expired packets and source symbols from its state. However, \shortn{} requires that the SIDs also monotonically increase as clients reset all streams below the value given in the \texttt{MC\_EXPIRE} frame. An improvement would replace this value by ranges of expired streams, but it would consume more bytes. 

\textbf{Congestion and flow control.} Multicast congestion control has been addressed by different researchers~\cite{rizzo2000pgmcc, widmer2001extending, matrawy2003survey} with attempts to build general-purpose TCP-friendly~\cite{rizzo2000pgmcc} congestion controllers or multicast application-oriented algorithms~\cite{widmer2001extending}. We leave the design of a multicast-specific congestion controller for \longn{} as a future work. 
Multicast applications are usually self-tailored by their nature. Live events are limited in their bandwidth consumption. Large software updates (i.e., file transfers) are generally performed in background. Leveraging the \emph{multicast as a service} paradigm, a client experiencing congestion could fall back on its unicast session for data delivery.
Flow control is also disabled for \shortn{} communication. In practice, limits like the maximum number of streams are set to maximum values.



\subsection{Source authentication}\label{sec:authentication}
Applications can decide to authenticate the source of their packets~\cite{canetti1999multicast,judge2003security}. This mechanism offers additional protection for use cases where malicious hosts could send spoofed packets, at the cost of more processing on the source and receivers.
For multicast, a natural method consists in adding an asymmetric signature to the packet sent. 
This signature can be verified by all clients on the multicast channel as only the intended source has access to the private key.
However, asymmetric signatures are computationally costly to generate and to verify on end-hosts.

\textbf{Overview.} \shortn{} supports two different source authentication methods whose performance depends on the group size. The main objective is to keep the additional authentication processing on the source as independent as possible of the group size.
The first method uses asymmetric signatures concatenated at the end of the packets sent.
In the second method, the multicast source sends an authentication packet on a second multicast channel containing symmetric authentication tags of the data packets using the unicast session keys of each client individually.

The \texttt{MC\_ANNOUNCE} frame sent by the \emph{UC Server} (Section~\ref{sec:design-channel}) additionally contains
\begin{enumerate*}[label=(\roman*)]
    \item the path authentication type (\emph{auth type} in short) and
    \item a path authentication type-specific payload (\emph{auth type payload}).
\end{enumerate*}
The \longn{} source can decide to disable source authentication.

\textbf{Signature.} By their nature, signatures are well suited for multicast source authentication. Even if the signature generation and verification is costly, it scales independently of the multicast group size. When signatures are used, the \emph{auth type payload} contains the public key used to verify the signature. The current implementation supports Ed25519~\cite{ed25519}, which produces signatures of 64 bytes. Other signature algorithms can easily be added, as long as the \emph{auth type payload} contains an additional field indicating the used algorithm and the signature length.

The signature is computed on the UDP payload containing (potentially multiple coalesced) QUIC packets. 
Clients verify the signature before decryption and processing of the QUIC packets.
The signature is concatenated at the end of the UDP datagram payload. The maximum size for QUIC packets is then reduced by the signature length, which is known by the client. 

\textbf{Authentication tags.} Computing symmetric authentication tags is less demanding than signatures, especially on short messages. This method relies on secret keys derived between a client and its \emph{UC Server} ($K_{1}$ and $K_2$ in \figurename~\ref{fig:design:multipath}). 

\begin{figure}
    \centering
    \includegraphics[scale=0.7]{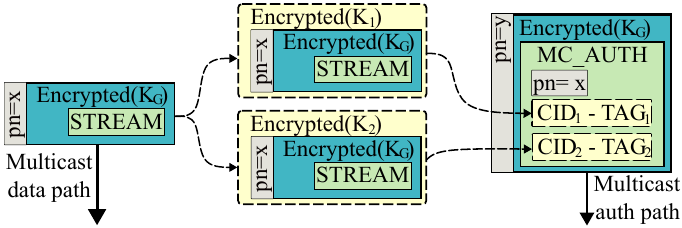}
    \caption{Symmetric source authentication design.}
    \label{fig:symmetric-design}
    \vspace{-1em}
\end{figure}


\figurename~\ref{fig:symmetric-design} presents the design of this method.
For each packet sent on the multicast path, the \emph{MC source} computes a symmetric authentication tag for each client, using the unicast session key shared with this client. The authentication payload is the encrypted packet sent on the multicast path. In our first implementation, we do not use a message authentication code (MAC), due to the lack of supported implementation in the cryptographic libraries we use. Instead, we use an authenticated encryption with associated data mechanism (AEAD) that explicitly outputs an authentication tag, namely AES in GCM mode. We ignore the ciphertext part of the output, and keep the 
tag, leaving the IV implicit as it was computed during the unicast QUIC handshake with the client. In effect, we use this AEAD to re-encrypt, using the unicast session keys, the encryption of each packet that is sent on the multicast path.
The resulting authentication tags are added in an \texttt{MC\_AUTH} frame (\tablename~\ref{table:frame-auth}), along with the source Connection ID of the client whose unicast session key was used to generate the tag ($\text{CID}_1$ and $\text{CID}_2$ in the \figurename). Finally, the frame contains the packet number of the packet it authenticates. There may be multiple \texttt{MC\_AUTH} frames if there are too many clients in the group to fit in a single QUIC packet.
To make the distinction between data and authentication packets, the clients maintain a second multicast path. 
We make the further distinction between the \emph{multicast data path} and the \emph{multicast auth path}, respectively for transporting multicast data and the \texttt{MC\_AUTH} frames.
Both paths use the same multicast address, but with different \emph{Channel ID}s and UDP destination ports.
The packets sent on the \emph{multicast auth path} are also encrypted with $K_G$. 
Clients map the received packets on the \emph{multicast data path} with the corresponding \texttt{MC\_AUTH} tag by decrypting the packet header to retrieve its packet number.

Clients may buffer packets received on the \emph{data path} until they receive the corresponding authentication information on the \emph{auth path}, 
or decide to directly process its payload without verifying the source. 
The client decrypts any incoming packet from the \emph{auth path} with $K_G$. It then looks for its Connection ID in the \texttt{MC\_AUTH} frame and retrieves the AEAD tag and the target packet number. 
If the retrieved packet number corresponds to a buffered (or already processed) packet received on the \emph{multicast data path}, the client encrypts it using its unicast session key. 
If the obtained tag is similar to the tag from the \texttt{MC\_AUTH} frame, the packet comes from the correct source and can safely be processed. As other clients of the group do not have access to the unicast session key, it is not possible for an adversarial member to send authentication packets with a valid AEAD tag to the group. Unauthenticated packets may be buffered by the client application until their expiration (triggered by an \texttt{MC\_EXPIRE} frame).


\textbf{Optimizations.} We detail optimizations speeding up the authentication process and decreasing the byte overhead.
First, adding the Source Connection ID of each client alongside the generated AEAD tag may be costly. Instead, our implementation generates for each client a \emph{Client ID}, a monotonically increasing unique client identifier. This value is generated by the \emph{MC source} and sent in the \texttt{MC\_KEY} frame when the client joins the multicast channel. 
The Client ID is encoded in 64 bits, less than the source CID.

Second, the method performs too many encryptions. The QUIC encryption procedure first encrypts the packet payload and generates the associated tag with the same key (in this context, the multicast group key, $K_G$). On the other hand, the authentication tags that are sent inside the \texttt{MC\_AUTH} frames are generated after the re-encryption of the data packets already encrypted with $K_G$. This new encryption is costly but of no utility as only the tag will be forwarded to the client in the \texttt{MC\_AUTH} frame. An improvement to this method would be to decouple the packet payload encryption from the tag authentication. The encryption would still be performed with $K_G$. The authentication would be repeated for each client with its corresponding unicast session key. As such, for $n$ clients, only a single encryption and $n$ tags would be executed. The current method computes $n+1$ encryption and $n+1$ tags.

\textbf{Authenticate groups of packets.} The two aforementioned methods authenticate the source at the packet level. We extended the asymmetric authentication method to enable the source to authenticate chunks of application data, potentially spanning several packets. Inside the QUIC packet carrying the end of an authenticated chunk of data, the source adds an \texttt{MC\_ASYM} frame (\tablename~\ref{table:frame-asym}). This frame embeds the asymmetric signature computed on the whole authenticated data using the asymmetric private key of the source. A receiver can authenticate each chunk of data it receives by verifying the signature on the received stream of data.

This method requires the recipient to wait for the entire chunk of data to be received to authenticate it as a whole. The application is responsible to cut its data in appropriate pieces. For example, a video conference application could authenticate each video frame separately.
In \shortn{}, chunks of data are mapped into streams. Each stream is hence signed individually, and the \texttt{MC\_ASYM} frame is sent after the last \texttt{STREAM} frame of the stream it authenticates. We further refer to this method as \emph{per-stream} asymmetric source authentication.


\subsection{Related work}\label{sec:design-ietf}

Two multicast QUIC approaches have been discussed within the IETF.
HTTP over multicast QUIC~\cite{pardue-quic-http-mcast-11} is an IETF draft designing HTTP communication over QUIC leveraging an IP multicast network. The multicast behavior (such as channel discovery) is implemented using HTTP and not QUIC. 
This idea has been implemented in \texttt{nghq}~\cite{nghq}. The draft also suggests using Forward Erasure Correction to recover lost packets and digital signatures to authenticate the source. However, this project is not maintained as it only partially supports a 3-years old version of the draft~\cite{pardue-quic-http-mcast-11} and does not implement QUIC as defined in RFC9000~\cite{rfc9000}.


Multicast extensions for QUIC~\cite{jholland-quic-multicast-02} is another draft related to this paper. It specifies extensions inside QUIC to support multicast. Our design follows some guidelines from the draft, e.g., the \texttt{MC\_ANNOUNCE}, \texttt{MC\_KEY} and \texttt{MC\_STATE} of \shortn{} have similar behavior than this IETF document. 
However, our design suggests a more scalable approach when faced to a large audience. First, the draft~\cite{jholland-quic-multicast-02} suggests to only retransmit lost frames using the unicast session between the server and the client.
\shortn{} recovers lost packets with Forward Erasure Correction frames sent on the multicast channel, thus enabling distinct clients to recover from different losses using a single additional multicast packet.
Second, Multicast extensions for QUIC~\cite{jholland-quic-multicast-02} recommends a different approach where data packets sent on the multicast channel are authenticated by hashing them and sending them either individually on each client or on another multicast channel whose packets carrying these hashes are also authenticated using the same authentication mechanism. Such approach, at some point, requires packet hashes to be regularly sent on the unicast session of each client of the multicast group, which is not scalable for large multicast groups. It introduces a linear dependency between the number of recipients and the amount of bytes being sent, which is against the multicast philosophy. \shortn{} suggests different approaches to solve the source authentication problem, discussed in Section~\ref{sec:authentication}. Some of these methods scale independently of the number of clients listening to the multicast channel. 
Finally, \shortn{} leverages Multipath QUIC~\cite{ietf-quic-multipath-04}, which renders the communication with the clients more straightforward as the multicast channel is considered as a second path with a different cryptographic context. It thus offers seamless migration between unicast and multicast in a single protocol. 
We believe that our approach is more scalable both for the recovery and authentication mechanism. We could not compare the performance of \shortn{} with the draft~\cite{jholland-quic-multicast-02} as we could not find any up-to-date implementation.

The literature contains various other approaches that aim at amortizing the cost of computing a full signature per packet -- see \cite[Sec.~14]{BS23} for an introduction. We do not rely on these techniques, as they introduce an important additional complexity and are not supported by standard libraries. The TESLA protocol~\cite{tesla} introduces a clever way to replace each signature with a single MAC, independently of the number of clients, but at the cost of introducing a strong constraint on the timing of packets, which may not be desirable in many applications. 

\section{\shortn{} performance benchmarks}\label{sec:benchmarks}

Real implementations of protocols are important to drive research as they prove the feasibility of an architecture.
Evaluating these implementations allows measuring their limits and behaviors in particular situations.

 An important contribution of this paper is a real implementation of \shortn{}. We extended the Cloudflare's \emph{quiche} open source project~\cite{quiche} with the design presented in Section~\ref{sec:design}. We leveraged pull request \#1310~\cite{quiche-multipath} providing support for Multipath QUIC as defined by the IETF draft~\cite{ietf-quic-multipath-04}. Our extension consists in $\sim$7000 Source Lines of Code (SLoC) for the design and $\sim$5000 SloC for tests. 
We reuse the implementation of the Forward Erasure Correction recovery mechanism from previous work~\cite{squirl}. The FEC code leverages Vandermonde matrices~\cite{klinger1967vandermonde} to generate coefficients used to generate the repair symbols, allowing multiple repair symbols to be computed with the same set of source symbols.
This section benchmarks the key features of \shortn{} in \emph{quiche}.

Jaeger \emph{et al.} already evaluated the performance of several QUIC implementations~\cite{jaeger2023quic}, including \emph{quiche}. Their measurement framework starts a client and a server on two machines connected with a \SI{10}{Gbps} link. In these experiments, the entire network stack is part of the evaluation. Even if the results present performance differences between implementations and show their speed on the wild~\cite{jaeger2023quic}, they do not isolate the QUIC processing. The true performance of a QUIC implementation also heavily depends on the network stack used to send UDP packets. For example, Tyunyayev \emph{et al.}~\cite{tyunyayev2022high} used kernel bypass techniques to improve QUIC throughput in a single connection by a factor of 3 without modifying the QUIC packet processing.

\shortn{} implements several algorithms that are CPU intensive, including Forward Erasure Correction and source authentication. Instead of evaluating the entire network stack where QUIC is a single component, we evaluate QUIC separately. This is possible as QUIC is a user-space protocol which can be tested as a library. Applications ask the implementation to send specific data. QUIC then generates packets and sends them to the wire, either using a socket or with kernel bypass techniques~\cite{tyunyayev2022high}. 
The benchmarks in this Section isolate QUIC packet generation (Server) and processing (Client). \figurename~\ref{fig:bench-setup} summarizes this setup. A benchmark consists in asking \emph{quiche} to send a stream data (\texttt{stream\_send}). We measure the time required by the Server to generate the corresponding packets (calls to \texttt{send}). For the Client benchmark, we store (in their order of generation) the packets generated by the Server, and measure the time required for a \emph{quiche} Client to process them (calls to \texttt{recv}).

Our baseline is a unicast QUIC Server with a single Client without losses. We limit the maximum QUIC packet size to 1350 bytes. The Server generates QUIC packets corresponding to a \SI{100}{\mega\byte} stream at a goodput rate of \SI{6.77}{Gbps}, and the Client processes them at goodput rate of \SI{5.23}{Gbps}. 
As we isolate QUIC processing from the network stack, raw performances are not especially representative. 
Instead, we show relative goodput, i.e., the amount of application data that can be sent in a unit of time, with respect to the baseline. 
Benchmarks are executed on a server equipped with two 10-core Intel Xeon CPU E5-2687W v3 @\SI{3.10}{\giga\hertz}. We repeat each benchmark 100 times and report the median and standard deviation for each point.

\subsection{Forward Erasure Correction recovery}\label{sec:benchmark-fec}
We first evaluate the impact of the FEC recovery mechanism. Upon reception of an \texttt{MC\_NACK} frame, the Server may generate repair symbols and send them in new packets, slowing down the goodput. We evaluate this cost when increasing the number of \texttt{MC\_NACK} received by the Source.

\begin{figure}
\centering
\begin{minipage}{.48\textwidth}
  \centering
  \resizebox{.95\linewidth}{!}{\tikzset{
	mymat/.style={
		matrix of nodes,
		text height=2.5ex,
		text depth=0.75ex,
		text width=11.25ex,
		align=center,
		row sep=-\pgflinewidth
	},
}

\begin{tikzpicture}[>=latex]
	\coordinate (bottom left) at (-6.5,-2.7);
	\coordinate (top right) at (-2.5, 2);
	\matrix[mymat,style={nodes=draw}]
	(buffer)
	{
		\ldots\\
		pn=$x$\\
		pn=$x+1$\\
		pn=$x+2$\\
		\ldots\\
	};
	\node[draw=black, rectangle, minimum width=2cm, minimum height=1cm, above left=-1.5cm and 0.5cm of buffer] (app1) {Application};
	\node[draw=black, rectangle, below=of app1, minimum width=2cm, minimum height=1cm] (quic1) {QUIC};
	\node[draw=black, rectangle, below=1.5cm of quic1, minimum width=2cm, minimum height=1cm] (udp1) {UDP};
	
	\node[draw=black, rectangle, minimum width=2cm, minimum height=1cm, above right=-1.5cm and 0.5cm of buffer] (app2) {Application};
	\node[draw=black, rectangle, below=of app2, minimum width=2cm, minimum height=1cm] (quic2) {QUIC};
	\node[draw=black, rectangle, below=1.5cm of quic2, minimum width=2cm, minimum height=1cm] (udp2) {UDP};
	\path let \p1 = ($(quic1.east)!0.7!(udp1.east) $ ) in node  at (-4.5,\y1) (center1) {};
	\path let \p1 = ($(quic2.east)!0.7!(udp2.east) $ ) in node  at (6,\y1) (center2) {};
	\draw[color=mypurple,thick] (center1) -- (center2);
	\node[text=mypurple] (userspace) at (4.8, -2.5) {User-space};
	\node[text=mypurple] (userspace) at (4.8, -3.1) {Kernel-space};
	
	\draw[->, very thick,myorange]  (quic1.south) -- ++(0,-0.75cm) node[pos=0.25, xshift=3.25ex,yshift=-2ex] {\texttt{send}} -| (buffer.south);
	\draw[<-, very thick,myorange] (quic2.south) -- ++(0,-0.75cm) node[pos=0.25, xshift=3.25ex,yshift=-2ex] {\texttt{recv}} -| ++(-1.5cm,0) -| ++ (0, 4.5cm) -| (buffer.north);
	
	\draw[->, very thick,mygreen] (app1.210) to node [xshift=7ex] {\texttt{stream\_send}} (quic1.150);
	\draw[<-, very thick,mygreen] (app2.330) to node [xshift=-7ex] {\texttt{stream\_recv}} (quic2.30);
	
	\coordinate (bottom left) at (-3.75,-2.5);
	\coordinate (top right) at (-1.5, 2.25);
	\node[draw=myblue, thick, text=myblue, rectangle, dashed,fit={(bottom left) (top right)}, align=right, text height=1ex] {Server Benchmark};
	
	\coordinate (bottom left2) at (1.5,-2.5);
	\coordinate (top right2) at (3.75, 2.25);
	\node[draw=myblue, thick, text=myblue, rectangle, dashed,fit={(bottom left2) (top right2)}, align=right, text height=1ex] {Client Benchmark};
	
\end{tikzpicture}}
  \captionof{figure}{Benchmark setup.}
  \label{fig:bench-setup}
\end{minipage}
\hfill
\begin{minipage}{.5\textwidth}
  \centering
  \includegraphics[width=\linewidth]{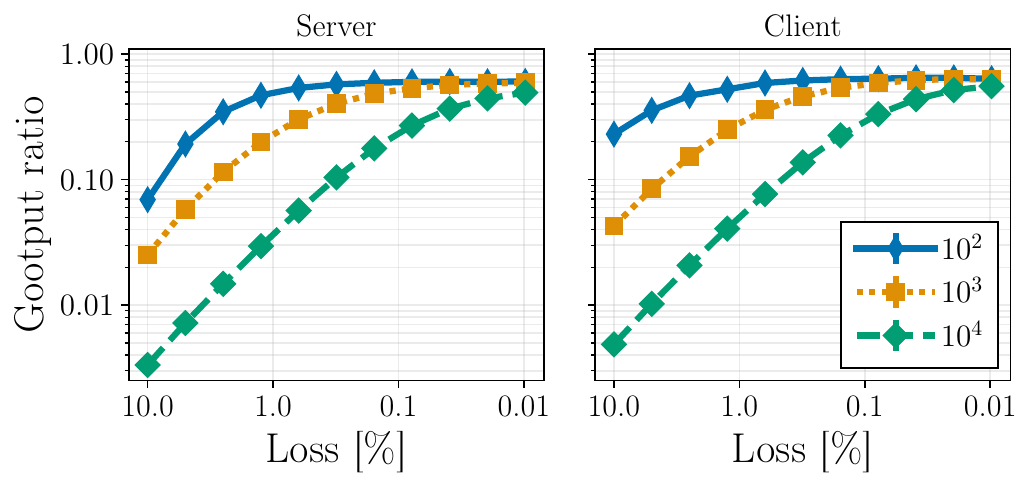}
  \captionof{figure}{Benchmark of the recovery mechanism using FEC with an increasing number of losses and FEC window sizes of 100, 1000 and 10000 symbols.}
  \label{fig:bench-repair}
  \vspace{-1.5em}
\end{minipage}
\end{figure}

\textbf{Benchmark setup.} We configure the multicast source to send a stream of \SI{100}{\mega\byte}.
Packet losses are simulated by notifying the multicast source with an \texttt{MC\_NACK} frame every $x$ packets ($1/x$~\% loss). The multicast source generates as many repair symbols as required to cope with these NACKs. We disable source authentication and set an expiration timer long enough to avoid expiring packets. 
\figurename~\ref{fig:bench-repair} presents the results.
We repeat the experiment with three different FEC window sizes, i.e., the maximum number of source symbols that are protected by a generated repair symbol. 
In practice, the FEC window size depends on the expiration timer, but we fix it to different values to see its impact on performance.

\textbf{Server and Client benchmarks.} As expected, increasing the FEC window size decreases the application goodput, as more protected symbols are used to generate a new repair symbol. This trade-off between higher performance (i.e., a smaller window) and higher protection (i.e., a larger window) is function of the expiration timer (ET) set by the application. A longer ET means 
that source symbols remain longer in the FEC window. 

Increasing the number of losses decreases performances. Below \SI{1}{\percent} losses, the performance reaches a plateau at $\sim$\SI{60}{\percent} of the baseline. This overhead comes from the implementation of the mapping from \texttt{STREAM} frames to source symbols. To avoid modifying the code of \emph{quiche} too much, we manually copy each protected-frame inside the FEC source symbol window. A zero-copy implementation would be possible at the cost of deeper changes in the core of the \emph{quiche} packet generation. Moreover, a \texttt{SOURCE\_SYMBOL} frame is added for each FEC-protected frame, decreasing by up to \SI{40}{\byte} the space available for \texttt{STREAM} data in a QUIC packet.


\subsection{Scalability to large groups}\label{sec:benchmark-scalability}
We compare the maximum application goodput that \shortn{} reaches compared to unicast QUIC with an increasing number of receivers. We also analyze the impact of adding source authentication to the communication, as discussed in Section~\ref{sec:authentication}.

\textbf{Benchmark setup.} We increase the number of receivers, $n$, from 1 to 40, and measure the goodput to send a stream of \SI{100}{\mega\byte}. In the unicast ($UC$) case, the Server must send a copy of each packet to each of the $n$ clients. As the impact of the Forward Erasure Correction recovery mechanism has already been evaluated in Section~\ref{sec:benchmark-fec}, we disable FEC for the multicast evaluations, i.e., sent \texttt{STREAM} frames are not mapped into source symbols (Appendix~\ref{sec:appendix-bench-scalability} details how the FEC recovery mechanism impacts this benchmark).
We do not add losses in this benchmark.
To avoid being limited by the congestion window in the unicast case and be fair compared to \shortn{}, we also disable the congestion control algorithm for unicast benchmarks.


\begin{figure*}
\centering
\begin{minipage}{.5\textwidth}
  \centering
  \includegraphics[width=\linewidth]{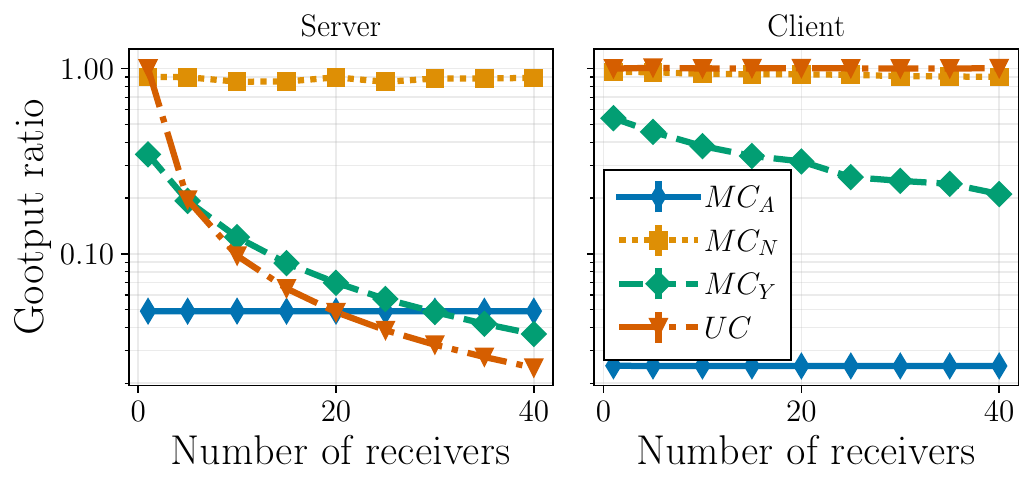}
  \captionof{figure}{Source authentication benchmark.}
  \label{fig:bench-source-auth}
\end{minipage}%
\begin{minipage}{.5\textwidth}
  \centering
  \includegraphics[width=\linewidth]{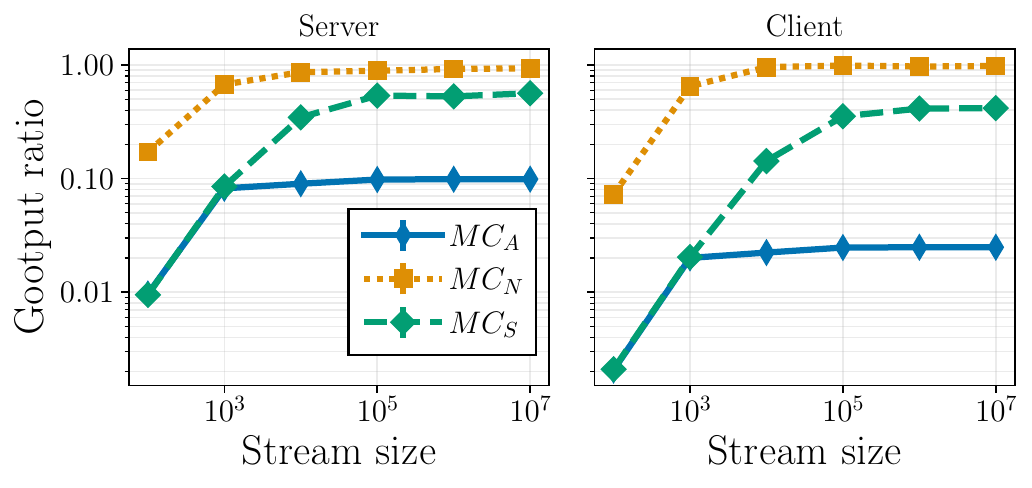}
  \captionof{figure}{Impact of stream size on $MC_A$ and $MC_S$.}
  \label{fig:bench-asym}
\end{minipage}
\end{figure*}

\textbf{Server benchmark.}
\figurename~\ref{fig:bench-source-auth}a presents the goodput ratio for the Server.
We first analyze \shortn{} without authentication ($MC_N$) with respect to unicast ($UC$). We see that for a single receiver, multicast and unicast achieve approximately same goodput. The multicast extension does not add significant overhead to the packet processing. However, by increasing $n$, we see that the unicast goodput (linearly) decreases with the number of receivers, but remains constant with $n$ for $MC_N$. As expected, the \shortn{} Server scales independently of the number of receivers.

Adding source authentication to the multicast delivery adds security at the cost of lower goodput.
\shortn{} with (asymmetric) signatures ($MC_A$) offers $\sim$20 times lower goodput than unauthenticated multicast. However, this method also scales with the multicast group size. For $n>20$, the cost of generating signatures is even less computationally intensive than unicast QUIC. For medium to large multicast communications, $MC_A$ can be used to offer source authentication and be more scalable for the source, without even considering the reduced consumed bandwidth.


\shortn{} with symmetric authentication ($MC_Y$) has an interesting behavior. For $n=1$, it offers approximately a third of the goodput of unicast. For each QUIC packet sent on the \emph{data path}, the method must generate an additional QUIC authentication packet. This packet contains the authentication tag computed using the symmetric key of the unicast session between the source and the unique receiver. The tag is computed by encrypting the QUIC data packet. The method thus computes three symmetric encryptions. Moreover, additional packet copies are required to perform these additional encryptions, adding overhead in the process. However, $MC_Y$ scales better with $n$, even outperforming unicast for $n>5$. Indeed, even if the number of packet encryptions is at least $n+2$ for this authentication method (instead of $n$ for unicast), fewer packets must be generated with $MC_Y$, thus reducing the processing cost. The goodput still decreases by increasing $n$ as more symmetric authentication tags must be computed and added inside authentication packets.

Finally, we observe that $MC_A$ offers higher goodput than $MC_Y$ for $n>30$. The cost of computing and adding symmetric authentication tags becomes higher than using an asymmetric signature on each data packet. An ideal approach for communications requiring source authentication could use the symmetric approach for $n<30$ and would switch to the asymmetric method for $n\geq 30$.

\textbf{Client benchmark.} Similarly, we analyze the impact of \shortn{} and its authentication methods on the achievable reception goodput on clients on \figurename~\ref{fig:bench-source-auth}b. First, we see that only the reception goodput of $MC_Y$ is impacted by $n$. The \texttt{MC\_AUTH} frames contain a Client ID/authentication tag pair for each multicast client. With higher values of $n$, a client must first decrypt longer packets as the \texttt{MC\_AUTH} frames carry more information. Then, it must look for its corresponding authentication tag in a longer list, increasing the total processing time. Fortunately, this processing overhead is less noticeable than the performance drop from \figurename~\ref{fig:bench-source-auth}a. For $n=1$, the symmetric authentication offers $~$\SI{60}{\percent} of the receiving goodput compared to unicast. Once more, this is mainly due to 
\begin{enumerate*}[label=(\roman*)]
    \item the decryption of the additional packet in the \emph{multicast auth} path for each \emph{multicast data} packet received and
    \item the encryption of the \emph{multicast data} packet to locally create the symmetric authentication tag to match the tag received in the \texttt{MC\_AUTH} frame.
\end{enumerate*}

The implementation without source authentication ($MC_N$) does not add significant overhead compared to unicast as both methods offer the same reception goodput. However, $MC_A$ has a strong negative impact on performance as the client must verify an Ed25519 signature for each packet. 
Compared to \figurename~\ref{fig:bench-source-auth}a, the bottleneck becomes the multicast client. A first option to improve the performances would be to choose a different asymmetric algorithm that decreases the overhead of signature verification, or use batch signature verification methods. A second option would be to offload signature verification to a dedicated thread. In that case, unauthenticated QUIC packets may be processed during signature verification, and dropped in case the signature is invalid. This horizontal scaling could be extended to several threads in a controller/worker fashion. Finally, dedicated hardware could speed up the signature verification.

\textbf{Stream-level authentication.} We now evaluate the performance cost of the \emph{per-stream} authentication mechanism. \figurename~\ref{fig:bench-asym} presents the goodput of \emph{packet-level} signatures ($MC_A$) and \emph{per-stream} signatures ($MC_S$). We vary the application stream size as it impacts the frequency of signature computation for $MC_S$. Again, we configure the multicast source to send \SI{100}{MB} of data. With a stream size of $c$ bytes, we expect $\frac{10^7}{c}$ streams to be sent.


We see the benefits of $MC_S$ over $MC_A$. There is a +\SI{400}{\percent} to +\SI{500}{\percent} improvement with the $MC_S$ method for streams of size $\geq 1000$ bytes. Concerning the receivers, we see an improvement of a factor between 8 and 10 for $\geq 1000$ bytes streams.
The performance of the unauthenticated method ($MC_N$) decreases for small stream sizes (\SI{100}{\byte} and \SI{1000}{\byte}). In practice, \emph{quiche} limits the number of different streams in a single QUIC packet to one. The Server hence sends a separate QUIC packet for each stream, increasing the number of packets compared to the baseline as these packets are shorter for the same amount of data to send. This adds overhead as more packets must be encrypted on the Server and decrypted on the Client.
For such stream sizes, the results are identical for $MC_S$ and $MC_A$ for the same reason.
As each stream fits in a single packet, the $MC_S$ method must create and send an \texttt{MC\_ASYM} frame in all multicast packets, needing as many signatures as $MC_A$.

We see a plateau for $MC_S$ for stream sizes $\geq 10^5$ B. We noticed that for such large sizes, the cost of hashing the entire stream for authentication becomes the limiting factor, balancing with the decreased number of required signatures. However, the performance of $MC_S$ is much closer to the baseline compared to $MC_A$. For $MC_A$, the plateau is reached earlier, as this method must authenticate each packet individually.

\section{\shortn{} Evaluation}\label{sec:evaluation}
This section performs end-to-end evaluations of \shortn{}.
We first show that \shortn{} can be deployed in a multicast-capable campus network. 
We then explore lossy scenarios with a larger set of receivers through emulations.
Throughout this Section, we consider two use cases for multicast communication, i.e., video conferencing\footnote{Such applications can be considered as one-to-many since they usually rely on a duplication server (DS). This DS receives video frames from each client and forward them to all the others.} and software updates through file transfers.

\textbf{Video conferencing.} We collected a trace of a five-minutes video call with Tixeo, a secure video conference application~\cite{tixeo}. The trace recorded the time at which Tixeo sent each video frame, as well as their sizes in byte. We consider that the first frame is sent at time 0.
Frame sizes vary from $\sim$\SI{1000}{\byte} to more than \SI{12}{\kilo\byte}. An experiment replays the trace. Each video frame is sent in a different QUIC stream. For simplicity, we consider that only the multicast source is sending data to the receivers, and that no client generates video frames that must be shared to the others.

Such application can tolerate few lost video frames but requires low latency.
We measure the \emph{frame lateness} on the clients for each sent video frame. It estimates the additional time required for a video frame to arrive at the clients independently of the initial delay from the network. It is computed as the difference between the time at which the whole video frame was received by the client, and the time at which the frame should have been sent by the source (i.e., when Tixeo sent the frame to the transport protocol, as indicated in the trace file). On each client, we further subtract the one-way delay between the client and server. This value is approximated by the minimum lateness computed on the client. The \emph{frame lateness} thus does not take into account the one-way delays between the source and the clients.

\textbf{File transfer.} Software updates can benefit from multicast. Multicast file transfers are usually performed in background at a relatively low bitrate to avoid causing congestion. We emulate this use case by sending \SI{1100}{\byte} payload packets at a fixed bitrate. Such use case requires that the same content is delivered reliably to all recipients, but is rather flexible on the reception latency. We measure the ratio of clients that receive all the content without error with \shortn{}. We show that even if \shortn{} is partially-reliable, its recovery mechanism can support heavy losses without impact on the receivers with a relatively low expiration timer.

\subsection{Experiments in a campus network}\label{sec:real-network}

\begin{figure*}
    \centering
    \includegraphics[width=\linewidth]{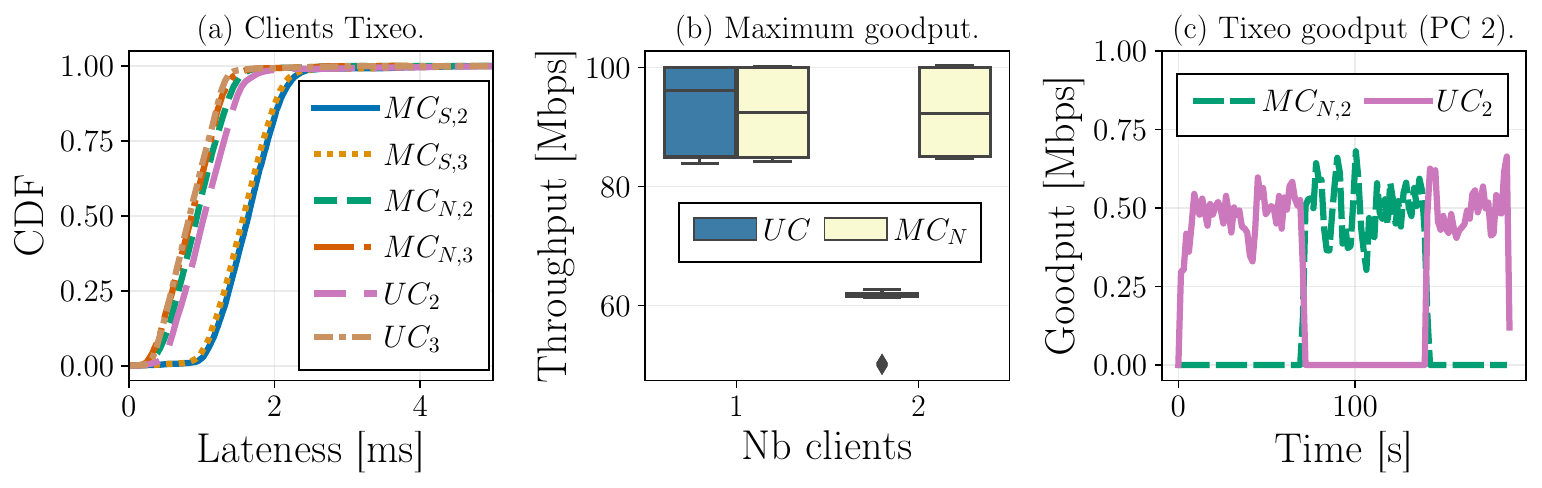}
    \caption{Results of \shortn{} deployed in our campus network, using PIM-SM. The integer suffix in \figurename~\ref{fig:didactique}a and \figurename~\ref{fig:didactique}c indicates the PC identifier.}
    \label{fig:didactique}
    \vspace{-1em}
\end{figure*}

We deploy \shortn{} in our network composed of two campuses separated by $\sim$\SI{23}{\kilo\meter}. 
Our IT infrastructure already leverages multicast for some software updates. 
It uses PIM Sparse Mode (PIM-SM)~\cite{rfc7761} and the RendezVous Point (RP) is located in the main campus. 
This network only supports IPv4 multicast.
Even if PIM-SM supports Any Source Multicast, we limit experiments to a single source.
We connect on this internal network three Intel PCs~\cite{intel-nuc} with \SI{1.6}{\giga\hertz} Intel Pentium processors and \SI{8}{\giga\byte} RAM. PCs 1 and 2 are located in the main campus and PC 3 is on the secondary site. The PCs are connected to the network with \SI{100}{Mbps} links.
\tablename~\ref{table:nuc} in Appendix~\ref{sec:appendix-nuc} provides information regarding the locations of these PCs, the latencies between them and the RP, and the number of hops separating them. PC 1 is our multicast source and the two others act as listeners. \figurename~\ref{fig:didactique} presents the results in this setup.

\textbf{Video trace.}
We compare the frame lateness of unicast and multicast using the video communication trace. \figurename~\ref{fig:didactique}a shows the Tixeo frame lateness on the receivers.
As expected by the low number of receivers, the impact of multicast is not significant regarding the frame lateness. The \emph{per-stream} authentication method ($\text{MC}_{S}$) adds $\sim$\SI{0.5}{\milli\second} of lateness at the receivers due to the CPU cost of the verification of the asymmetric signatures. 
This small experiment demonstrates that our implementation of \shortn{} works in a real multicast network. 

We did not measure any loss with IP multicast in our campus network. Past research in the MBone~\cite{eriksson1994mbone}, the first large multicast network used for research, measured losses below \SI{3}{\percent}~\cite{caceres1999inference, yajnik1996packet} even if we can expect lower rates in today's multicast networks. Recent studies show losses on the Starlink medium~\cite{michel2022first} in the order of \SI{0.40}{\percent} while QUIC designers \cite{langley2017quic} reported TCP retransmission rates of \SI{1}{\percent}, \SI{2}{\percent} and \SI{8}{\percent}.

We also measured the maximum throughput on PCs 2 and 3 with \shortn{} compared to unicast QUIC using packets of \SI{1100}{\byte} payload on \figurename~\ref{fig:didactique}b. 
We do not use source authentication on multicast ($MC_N$).
For a single client, $MC_{N,2}$ provides similar throughput as unicast QUIC ($UC_{2}$). 
Increasing the audience to two receivers gives approximately the same performance for \shortn{}, while this value is almost halved for unicast. This small result shows that \shortn{} can saturate a \SI{100}{Mbps} link and offers scalability advantages compared to its unicast equivalent in a real network.

Finally, \figurename~\ref{fig:didactique}c shows that \shortn{} receivers can seamlessly switch from unicast to multicast while still receiving data. The source sends the Tixeo video trace. PC~3 immediately joins the multicast group. PC~2 starts as a unicast receiver and then joins the multicast group after 70 seconds by sending an \texttt{MC\_STATE} frame with the \texttt{JOIN} action to the source. PC~2 then receives application data through the multicast path ($MC_N$) instead of the unicast one ($UC$). 70 seconds later, PC~2 leaves the multicast channel and stops listening to the IP multicast address. The source sends the next data to PC~2 on the unicast path.

\subsection{Experiments in emulated networks}\label{sec:emulations}

We now explore different scenarios with more receivers under losses using Mininet~\cite{lantz2010network}. The emulated network is a binary tree whose root is connected to the multicast source, \emph{S}. Each node in the tree is a client of \emph{S}. The binary tree contains 50 clients, by filling up each layer of the tree before creating a new layer. Multicast routes are statically set from the root to each leaf with \texttt{smcrouted}. Each link has a bandwidth of  \SI{100}{Mbps} and a delay of \SI{3}{\milli\second}.
We run this network on the same server used for the benchmarks in Section~\ref{sec:benchmarks}.




\textbf{Video trace under losses.}
We add losses in the network using \texttt{tc}. The losses are added on the link between each leaf-node client and its parent in the multicast tree. As a result, only half the receivers are impacted by erased packets, and each lossy client experiences different losses than the others. We explore this scenario to estimate the performance of our Forward Erasure Correction recovery mechanism where a single repair symbol frame can recover distinct losses on different clients. 
An opposite scenario is to add the loss on a single link at the top of the tree; clients thus experience the same losses and this can lead to the well-known NACK-implosion problem. This scenario is explored in Appendix~\ref{sec:appendix-losses}.
As \shortn{} does not currently include a congestion controller, we also disable it for the unicast experiments.
We set the expiration timer to \SI{350}{\milli\second}. Above this value, the utility of a video frame becomes low due to the interactivity of the application. For example, Zoom recommends a latency below \SI{150}{\milli\second} for good quality of experience~\cite{zoom-latency}. 

\begin{figure*}
    \centering
    \includegraphics[width=\linewidth]{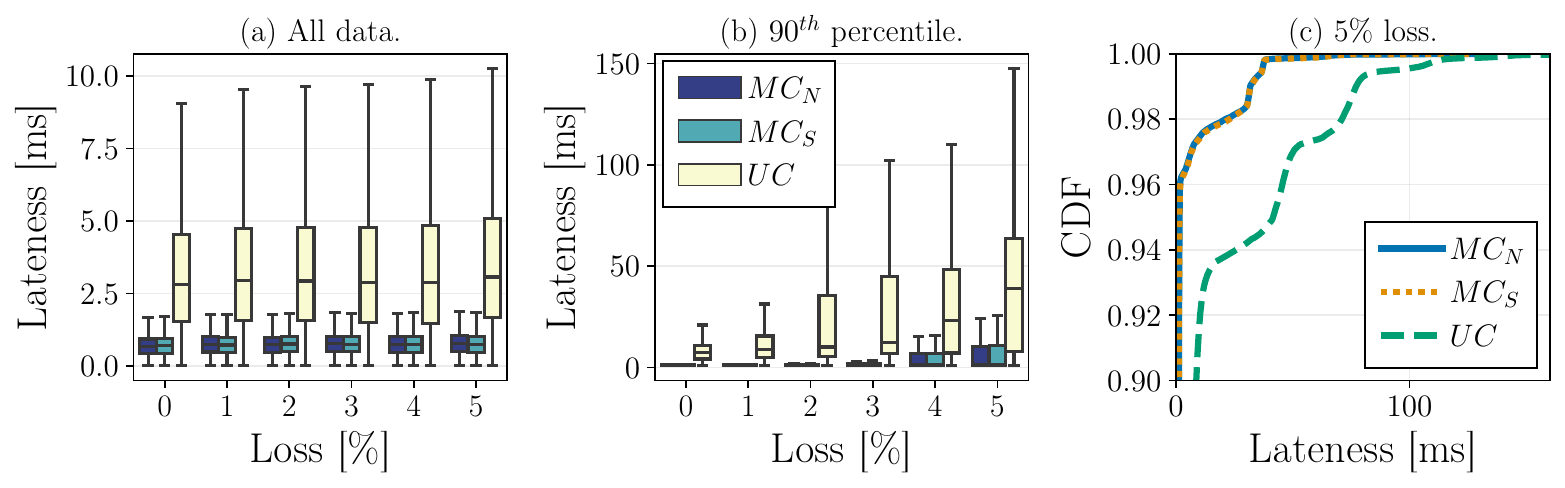}
    \caption{Impact of packet losses on the video frame lateness with 50 receivers. Each boxplot contains frame latenesses aggregated on all received frames from all receivers. \figurename~\ref{fig:losses}a and~\ref{fig:losses}b do not show outliers for readability. The boxes represent the \nth{25} and \nth{75} percentiles among the used data and the whiskers the 1.5 IQR. For \figurename~\ref{fig:losses}b, it means that the boxes contain the $92.5^{\text{th}}$ to $97.5^{\text{th}}$ percentile of the whole experiment results. \figurename~\ref{fig:losses}c highlights the cumulative distribution function for a loss of \SI{5}{\percent} for the last percentiles.}
    \label{fig:losses}
    \vspace{-1em}
\end{figure*}

\figurename~\ref{fig:losses}a shows the frame lateness aggregated on all received frames on all 50 clients for \shortn{} without source authentication ($MC_N$), with \emph{per-stream} authentication ($MC_S$) and using unicast ($UC$).
The frame lateness with multicast does not significantly increase with the added loss as these losses only impact a small percentage of the packets. We expect that the majority of the frame latenesses remain identical. 
However, the \figurename\ shows that the frame lateness on the clients is higher with unicast compared to multicast. Even without losses, the unicast source must individually send the same data to all clients, inducing a frame lateness increase on the clients.

\begin{figure*}
\begin{minipage}{.64\textwidth}
  \centering
  \includegraphics[width=\linewidth]{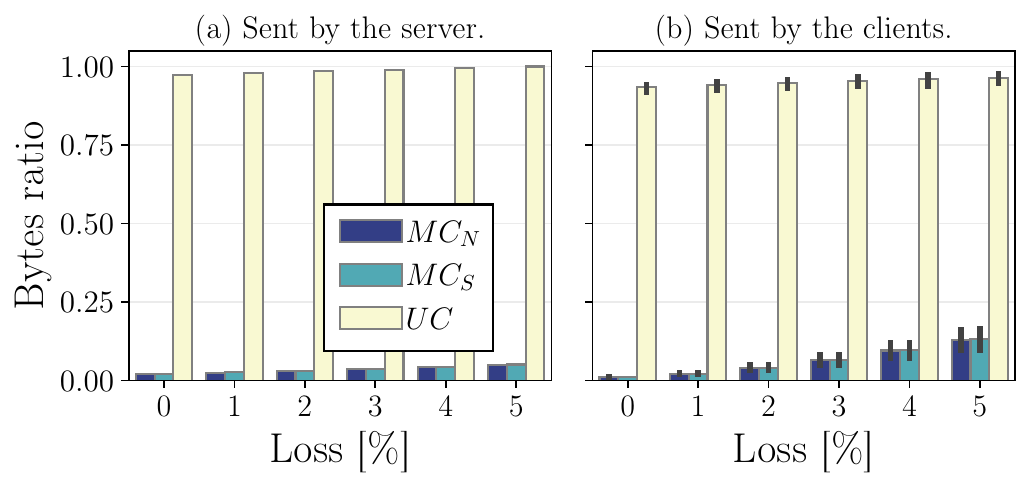}
  \captionof{figure}{Impact of packet losses on the number of UDP payload bytes sent by the source~(left) and all the receivers~(right), aggregating data sent on the unicast and multicast path.}
  \label{fig:losses-bytes}
\end{minipage}%
\hfill
\begin{minipage}{.33\textwidth}
  \centering
  \includegraphics[width=\linewidth]{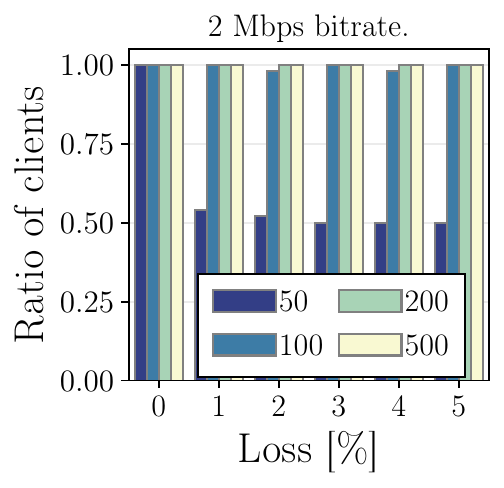}
  \captionof{figure}{Ratio of \shortn{} clients receiving all the content of the file transfer when varying the ET.}
  \label{fig:losses-files}
\end{minipage}%
\vspace{-1em}
\end{figure*}

\figurename~\ref{fig:losses}b only presents the $90^{\text{th}}$ percentile of frame latenesses. As expected, the last percentiles are more affected by packet losses since they require one or several RTTs to carry out the retransmission. The \figurename\ highlights the scalability of our FEC repair mechanism as a single FEC repair packet is sufficient to recover distinct losses on different clients. The frame lateness slightly increases with the loss percentage in the multicast scenario as an increasing number of packets are lost and must wait for a negative acknowledgment from clients to trigger repair packets. \figurename~\ref{fig:losses}c presents the data using a cumulative distribution function (CDF) for the \SI{5}{\percent} loss experiment.
Even in this severe scenario, \shortn{} outperforms unicast, both with \emph{per-stream} and without source authentication. We note that $\sim$\SI{96}{\percent} of the frames are not impacted by the losses with \shortn{}, whereas it reaches almost \SI{50}{\milli\second} for unicast communication. Such added frame lateness can have a negative impact for video conferences as the playback buffer allowed for the application decreases.

The main benefit of multicast comes from the reduction of the number of bytes sent in a network. \figurename~\ref{fig:losses-bytes} reports the number of bytes sent by the source (left) and the receivers (right). Without losses, unicast QUIC ($UC$) sends $\sim$50 times more bytes than multicast ($MC_N$ and $MC_S$) because it must replicate each QUIC packet to each of the 50 receivers. The bandwidth consumption increases almost similarly with \shortn{} compared to unicast QUIC when increasing the losses. As each client at the bottom of the tree is affected by different losses, a repair symbol sent by the source is not always sufficient to recover losses from the other clients, thus requiring to send more repair symbols. Moreover, the \shortn{} source must acknowledge the \texttt{MC\_NACK} frames received from its clients.
These \texttt{MC\_NACK} frames are only sent when a receiver sees a gap in the packet number sequence. When the network is free of losses, an \shortn{} client only sends bytes at the initialization and termination of the connection. This is confirmed by \figurename~\ref{fig:losses-bytes}b. When we increase the losses, clients send more \texttt{MC\_NACK} frames to the server. 


\textbf{File transfer under losses.}
All chunks of a software update must be received by all clients in order to be successful. We start a \SI{5}{\mega\byte} file transfer on the source using \shortn{} with \emph{per-stream} source authentication. We use the same emulated topology of 50 clients and also add losses on the leaves of the tree. The source chunks the transfered file in \SI{1100}{\byte} streams at a rate of \SI{2}{Mbps}. We vary the expiration timer (ET) from \SI{50}{\milli\second} to \SI{500}{\milli\second}. \figurename~\ref{fig:losses-files} shows that even a small expiration timer of \SI{200}{\milli\second} is sufficient to ensure correct reception of data packets under \SI{5}{\percent} losses.
A lost packet is detected by the client when it sees a gap in the packet number sequence. Thus, a lost packet cannot be detected before (at least) the reception of the subsequent packet. A higher bitrate means that this delay is shorter to detect losses. This delay is of \SI{5}{\milli\second} for a \SI{2}{Mbps} bitrate. Considering that the delay between the source and the clients at the bottom of the tree is \SI{36}{\milli\second}, an expiration timer of \SI{50}{\milli\second} is not always sufficient to allow the client to detect the loss, ask for recovery through an \texttt{MC\_NACK} frame and receive the \texttt{REPAIR} frame.
Even under more losses than what was measured more than twenty-years ago on the MBone~\cite{yajnik1996packet,caceres1999inference}, a reasonable ET of \SI{200}{\milli\second} is sufficient at the measured bitrate to ensure the reliable delivery of a file to all clients.

\section{Conclusion}\label{sec:conclusion}
With QUIC, a single transport protocol can support a variety of application requirements as shows by the existing support for reliable streams \cite{rfc9000} and datagrams \cite{rfc9221}. This paper goes one step further with \shortn{}, an extension to QUIC that enables this protocol to support both unicast and multicast. \shortn{} provides scalable recovery mechanism with Forward Erasure Correction and source authentication mechanisms independent of the number of members in the multicast group. Furthermore, by leveraging the path management features of Multipath QUIC~\cite{ietf-quic-multipath-04}, \shortn{} can support multicast and unicast receivers in the same session and allow them to seamlessly switch from both methods. \shortn{} is implemented in the open-source \emph{quiche}~\cite{quiche} project, has been demonstrated in our campus network and evaluated using benchmarks and emulated networks.



To encourage other researchers and application developers to test \shortn{} in wide area networks, we will release our implementation of \shortn{} upon upcoming publications. Our next steps include discussing the design of \shortn{} within the IETF, adding multicast congestion controls and use it inside real applications to better understand how and when receivers should switch from multicast to unicast. 


\newpage


\bibliographystyle{ACM-Reference-Format}
\bibliography{paper}

\newpage

\appendix

\section{Complement on the scalability to large groups benchmark}\label{sec:appendix-bench-scalability}
Section~\ref{sec:benchmark-scalability} benchmarks the goodput ratio of \shortn{} without source authentication ($MC_N$), with the symmetric ($MC_Y$) and asymmetric source authentication ($MC_A$) with respect to unicast QUIC ($UC$) when increasing the number of receivers.

\figurename~\ref{fig:bench-source-auth} showed that $MC_N$ provides similar goodput as $UC$, even for larger groups. This benchmark disabled the Forward Erasure Correction (FEC) recovery mechanism. Indeed, as discussed in Section~\ref{sec:benchmark-fec}, the current implementation of this mechanism in \emph{quiche} is not optimal as it performs too many copied of the protected frames. This choice was made to minimize the changes in \emph{quiche} and hence facilitate maintenance with the upstream project.

\begin{figure}[h]
\centering
\begin{minipage}{.5\textwidth}
  \centering
  \includegraphics[width=\linewidth]{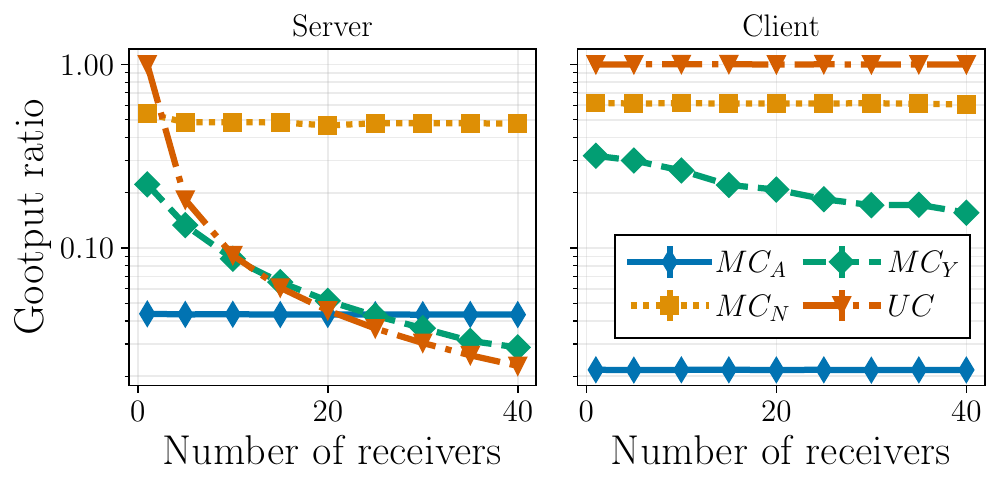}
  \captionof{figure}{Source authentication benchmark with the Forward Erasure Correction recovery mechanism enabled.}
  \label{fig:bench-source-auth-appendix}
\end{minipage}%
\end{figure}

However, this Appendix presents the same results as \figurename~\ref{fig:bench-source-auth} with the FEC recovery mechanism enabled. \figurename~\ref{fig:bench-source-auth-appendix} shows the results.

As expected (and highlighted in \figurename~\ref{fig:bench-repair}), enabling the FEC mechanism decreases the performance of \shortn{}. The $MC_N$ source now reaches \SI{50}{\percent} of the goodput baseline (unicast QUIC with a single client). However, the benefits of \shortn{} remain positive when the communication is not limited to a single client. Moreover, $MC_N$ and $MC_A$ performance is again independent of the number of receivers, $n$.
Compared to the results of \figurename~\ref{fig:bench-source-auth}, $MC_A$ outperforms $UC$ for $n>20$, and $MC_Y$ is better than $UC$ for $n>10$. Despite the small performance drop caused by the FEC recovery mechanism, the conclusion regarding the scalability of \shortn{}, with and without source authentication, remain identical to the discussion from Section~\ref{sec:benchmark-scalability}.

\section{\shortn{} deployment in the campus network}\label{sec:appendix-nuc}
Table~\ref{table:nuc} shows additional characteristics of the setup of the three NUCs in our campus network.

\begin{table*}[ht]
	\centering
	\begin{tabular}{ccccccc}
			\toprule
			& 1$\rightarrow$RP & 2$\rightarrow$RP & 3$\rightarrow$RP & 1$\leftrightarrow$2 & 1$\leftrightarrow$43& 2$\leftrightarrow$3 \\
			\midrule
			Lat. (ms) & 0.27 & 1.48 & 0.20 & 1.40 & 0.19 & 1.45 \\
			\midrule
			\# hops & 4 & 5 & 4 & 3 & 3 & 3 \\
			\bottomrule
		\end{tabular}
		\captionof{table}{Latency and number of hops of the three Intel PCs~\cite{intel-nuc} with the RendezVous Point (RP) and each other. PC 1, PC 4 and the RP are located in the main campus. PC 3 is located on the secondary campus.}
        \label{table:nuc}
        \vspace{-1em}
\end{table*}

\section{Video trace lateness with shared losses}\label{sec:appendix-losses}
\figurename~\ref{fig:losses} from Section~\ref{sec:emulations} presents the frame lateness using a trace from the video conferencing application Tixeo~\cite{tixeo}. It uses emulations with Mininet using a binary tree topology and emulates losses with \texttt{tc}. The losses are added on the links between each leaf node of the tree and its parent, causing potentially different losses on affected clients.

This Section evaluates the opposite scenario, where a single link is impacted with losses. We choose the link at the second layer of the tree to impact only \SI{50}{\percent} of clients. All affected clients are subjected to the same packet erasures. We use the same emulation setup as described in Section~\ref{sec:emulations}.

\begin{figure*}
    \centering
    \includegraphics[width=\linewidth]{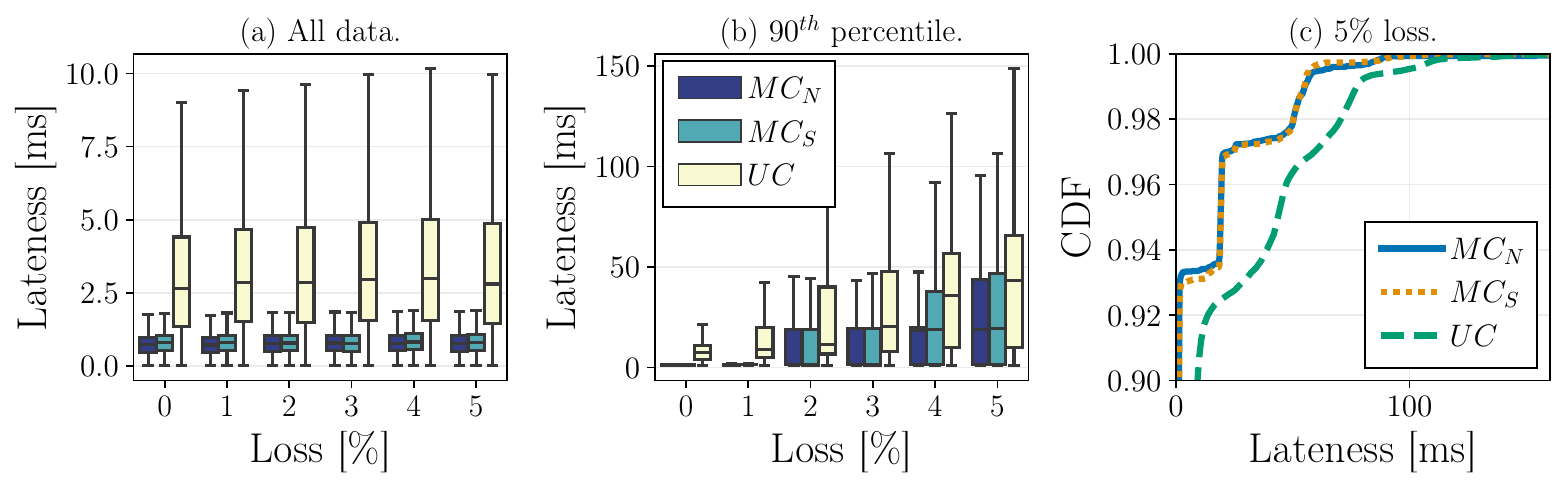}
    \caption{Impact of packet losses on the video frame lateness with 50 receivers. In opposition to \figurename~\ref{fig:losses}, all impacted receivers see the same packet erasures. The format of the \figurename\ is similar to \figurename~\ref{fig:losses}.}
    \label{fig:losses-appendix}
\end{figure*}

\begin{figure*}
\begin{minipage}{.66\textwidth}
  \centering
  \includegraphics[width=\linewidth]{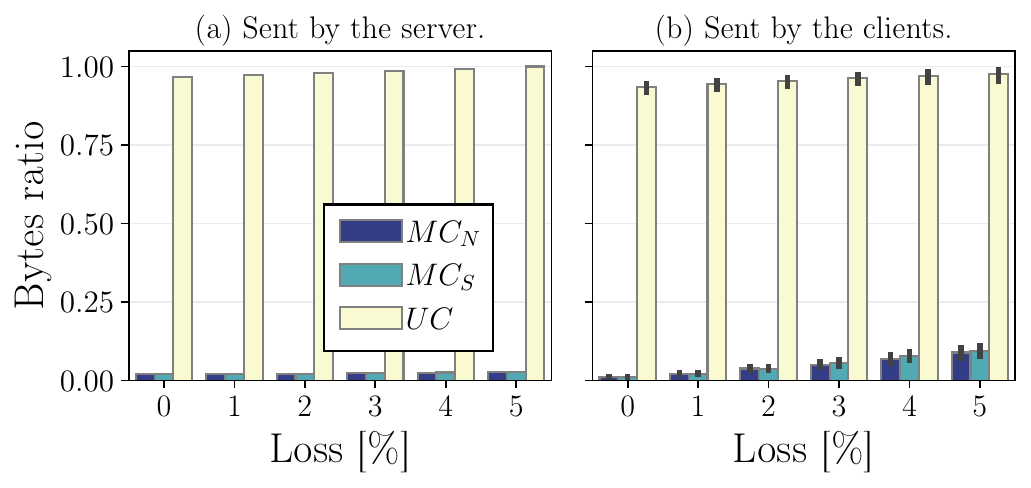}
  \captionof{figure}{Impact of packet losses on the number of UDP payload bytes sent by the source~(left) and the receivers~(right) when losses are added on the left branch at the top of the tree.}
  \label{fig:losses-bytes-appendix}
\end{minipage}%
\end{figure*}

The results are presented in \figurename~\ref{fig:losses-appendix}. Taking into account all data (\figurename~\ref{fig:losses-appendix}a) and only the last percentiles (\figurename~\ref{fig:losses-appendix}b), multicast still scales better than unicast even under losses. However, the frame lateness in this specific scenario is higher for multicast, compared to the results from \figurename~\ref{fig:losses}. 

In this scenario, the source is faced to the well-known NACK implosion problem. Even if negative acknowledgment (NACK) is more scalable than positive acknowledgment (e.g., used in unicast QUIC), a source may be faced to an implosion problem when a large subset of its receivers face the same packet losses. All recipients forward to the source, approximately at the same time (depending on their distance to the source) a NACK for the same packet. Even if this information is needed only once by the source, clients do not communicate with each other. Existing works~\cite{lehman1998active} suggest participation of the multicast network to avoid this NACK implosion problem by suppressing NACK duplicates before they reach the source. However, such methods require participation of the network. This would require sending non-encrypted \texttt{MC\_NACK} frames that any intermediate node could read in the network. Even if we believe that such methods may be part of \shortn{} in the future, we consider them as out of scope of this paper.

\figurename~\ref{fig:losses-bytes-appendix} shows the ratio of bytes sent by the source (left) and receivers (right) in this particular scenario. Compared to \figurename~\ref{fig:losses-bytes}, the increase in the bytes sent by the multicast source is lower when all clients are faced to the same losses. As the received \texttt{MC\_NACK} frames contain the same gaps in the packet number sequence, the source generates FEC repair packets only once, following the FEC scheduler from Listing~\ref{list:fec}.
Finally, \figurename~\ref{fig:losses-appendix} shows that even without NACK suppression, multicast performs better than unicast under severe losses. Furthermore, this problem could be alleviated by sending \emph{a priori} Forward Erasure Correction repair packets.

\section{\shortn{} frames}\label{sec:frames}
\begin{table*}[ht]
	\centering
	\begin{tabularx}{\textwidth}{lllX}
			\toprule
			Name & Type & Size (bytes) & Description \\
			\midrule
			\texttt{channel\_id} & Bytes & $\leq$20 & Identifies an \shortn{} path. \\
			\midrule
			\texttt{use\_ipv6} & Boolean & 1 & Whether the channel uses IPv6 (IPv4 is false). \\
            \midrule
            \texttt{source\_ip} & IP Addr & 4 or 16 & IP address of the source for SSM multicast. \\
            \midrule
            \texttt{group\_ip} & IP Addr & 4 or 16 & IP address of the SSM multicast group. \\
            \midrule
            \texttt{udp\_port} & u16 & 2 & UDP destination port of the SSM multicast group. \\
            \midrule
            \texttt{exp\_timer} & u64 & 8 & Expiration timer, in milliseconds. \\
            \midrule
            \texttt{auth\_type} & u64 & 8 & Source authentication method used. \\
            \midrule
            \texttt{path\_type} & u64 & 8 & Whether this \texttt{MC\_ANNOUNCE} frame announces a multicast data path or multicast authentication path. \\
            \midrule
            \texttt{pub\_key} & Bytes & Variable & Public key if the asymmetric method is used. \\
            
            \bottomrule
		\end{tabularx}
		\captionof{table}{\texttt{MC\_ANNOUNCE} frame. It announces multicast channel information to a client on the unicast connection.}
        \label{table:frame-announce}
        \vspace{-1.5em}
\end{table*}

\begin{table*}[ht]
	\centering
	\begin{tabularx}{\textwidth}{lllX}
			\toprule
			Name & Type & Size (bytes) & Description \\
			\midrule
			\texttt{channel\_id} & Bytes & $\leq$20 & Identifies the \shortn{} path. \\
			\midrule
            \texttt{action} & u64 & 8 & Action performed by the client, e.g., \texttt{JOIN}/\texttt{LEAVE}. \\
            \midrule
            \texttt{action\_data} & u64 & 8 & Additional action-specific data. \\    
            \bottomrule
		\end{tabularx}
		\captionof{table}{\texttt{MC\_STATE} frame. It carries state and actions performed by the \shortn{} client. It is sent on the unicast path.}
        \label{table:frame-state}
        \vspace{-1.5em}
\end{table*}

\begin{table*}[ht]
	\centering
	\begin{tabularx}{\textwidth}{lllX}
			\toprule
			Name & Type & Size (bytes) & Description \\
			\midrule
			\texttt{Signature} & Bytes & Variable & Asymmetric signature of the stream data it authenticates.\\
            \bottomrule
    \end{tabularx}
    \captionof{table}{\texttt{MC\_ASYM} frame. It follows the last \texttt{STREAM} frame of the stream it authenticates with asymmetric signature. It is sent on the multicast data path.}
    \label{table:frame-asym}
    \vspace{-1.5em}
\end{table*}

\begin{table*}[ht]
	\centering
	\begin{tabularx}{\textwidth}{lllX}
			\toprule
			Name & Type & Size (bytes) & Description \\
			\midrule
			\texttt{channel\_id} & Bytes & $\leq$20 & Identifies the \shortn{} path. \\
			\midrule
            \texttt{pn} & u64 & 8 & The highest packet number received by the client. \\
            \midrule
            \texttt{ranges} & Bytes & Variable & Range of packet numbers not received. \\
            \midrule
            \texttt{fec\_nack} & Bytes & Variable & FEC recovery-specific payload giving potentially additional information on the source symbols not received. \\
            \bottomrule
    \end{tabularx}
    \captionof{table}{\texttt{MC\_NACK} frame. It contains the gaps in the packet numbers on the client. It is sent by the client on the unicast path.}
    \label{table:frame-nack}
\end{table*}

\begin{table*}[ht]
	\centering
	\begin{tabularx}{\textwidth}{lllX}
			\toprule
			Name & Type & Size (bytes) & Description \\
			\midrule
			\texttt{channel\_id} & Bytes & $\leq$20 & Identifies the \shortn{} path. \\
			\midrule
            \texttt{key} & Bytes & Variable & Secret to generate the group key of the multicast path. \\
            \midrule
            \texttt{first\_pn} & u64 & 8 & The first packet number that the client can expect to receive. \\    
            \midrule
            \texttt{client\_id} & u64 & 8 & Unique identifier of the client for the server. \\    
            \bottomrule
		\end{tabularx}
        \captionof{table}{\texttt{MC\_KEY} frame. It carries the secrets to generate the group key of the multicast path. It is sent on the unicast path.}
        \label{table:frame-key}
        \vspace{-1.5em}
\end{table*}

\begin{table*}[ht]
	\centering
	\begin{tabularx}{\textwidth}{lllX}
			\toprule
			Name & Type & Size (bytes) & Description \\
			\midrule
			\texttt{channel\_id} & Bytes & $\leq$20 & Identifies the \shortn{} path. \\
			\midrule
            \texttt{exp\_type} & u8 & 1 & Map indicating which of the following fields are not empty. \\
            \midrule
            \texttt{pn} & u64 & 8 & Last expired packet number. All packets with a lower or equal packet number cannot trigger repair symbol generation anymore. \\    
            \midrule
            \texttt{stream\_id} & u64 & 8 & Last expired stream ID. All streams with a lower or equal stream ID cannot trigger repair symbol generation anymore. \\
            \midrule
            \texttt{fec\_metadata} & Bytes & Variable & FEC information about last expired source symbols. All source symbols expired through this field cannot trigger repair symbol generation anymore.\\
            \bottomrule
    \end{tabularx}
    \captionof{table}{\texttt{MC\_EXPIRE} frame. It notifies expired packets, streams and source symbols to the client. It is sent on the multicast path.}
    \label{table:frame-expire}
    \vspace{-1.5em}
\end{table*}

\begin{table*}[ht]
	\centering
	\begin{tabularx}{\textwidth}{lllX}
			\toprule
			Name & Type & Size (bytes) & Description \\
			\midrule
			\texttt{channel\_id} & Bytes & $\leq$20 & Identifies the \shortn{} path. \\
			\midrule
            \texttt{pn} & u64 & 8 & Packet number of the packet authenticated with this frame. \\    
            \midrule
            \texttt{signatures} & Bytes & Variable & For each client, a pair Client ID/symmetric signature.\\
            \bottomrule
    \end{tabularx}
    \captionof{table}{\texttt{MC\_AUTH} frame. It embeds symmetric signatures for each client of the packet it authentifies. It is sent on the multicast authentication path.}
    \label{table:frame-auth}
    \vspace{-1.5em}
\end{table*}

\end{document}